\documentclass[12pt]{iopart}
\input epsf
\begin{document}

\title{\bf Glassy Dynamics in a Frustrated Spin System:  Role of Defects}

\author{Bulbul Chakraborty, Lei Gu, Hui Yin}

\address{Martin Fisher School of Physics, Brandeis University, Waltham, MA
02254.}



\begin{abstract}
In an effort to understand the glass transition, the kinetics of a spin model
with frustration
but no quenched randomness has been analyzed. The phenomenology of the spin
model is remarkably similar to that of structural glasses.  Analysis of the
model suggests that defects play a major role in dictating the dynamics as the glass transition is
approached. 
\end{abstract}
\section{Introduction}
In recent years, there has been considerable interest in 
spin models without quenched-in disorder
which, however, exhibit glassy dynamics\cite{BouMez}. The model discussed in this paper belongs
to this category.  Frustration is a key concept in theories
of structural glasses such as the curved-space pictures of metallic
glass\cite{Egami}.  The two concepts taken together, glassy dynamics in
non-random spin systems and frustration as a key to glassy behavior, suggest
that frustrated spin systems could play a role in understanding
the nature of structural glasses\cite{Kivelson}. 

The compressible, triangular-lattice, Ising  antiferromagnet is an interesting
example of a system where the frustration is removed by elastic
distortions\cite{Chen1,Leigu1}.  The phase transition from a disordered,
paramagnetic phase to an ordered striped phase has been studied in
detail\cite{Chen1,Leigu1}, and in the last couple of years, interest has shifted 
to studying the dynamics of the supercooled state\cite{Leigu2,Ignatiev}.  Monte
Carlo simulations of the model\cite{Leigu2} have shown that following an
instantaneous quench from an initial  temperature $T_i >> T_F$ (the first-order transition 
temperature), to  a final temperature $T_f <
T_F$, the dynamics of the system changes character as $T_f$ is lowered below a
temperature $T^*$.  The observed changes are consistent with a glass
transition occurring at $T^*$.  As $T^*$ is approached from above, the characteristic
relaxation time increases dramatically, and the spin autocorrelation function
develops a plateau at intermediate times (Fig. \ref{fig1}).
Analysis of the fluctuation metric and the energy metric\cite{Mountain} shows that the
system becomes non-ergodic below $T^*$\cite{Leigu2}. 
Although, the dynamics in this non-ergodic phase is reminiscent of ``coarsening'',
the relaxation towards equilibrium is unusual and is 
characterized by long-time periods of quiescent behavior interrupted by rare
jump events. 
The distribution of jump-intervals obeys a power law\cite{Leigu2} which suggests
that the model belongs to the category of weak-ergodicity
breaking\cite{Bouch_weak}.

In this paper, the approach to the ``glass transition'' at $T^*$ is analyzed
through a study of the spin-spin autocorrelation functions and the behavior of
defects which arise naturally when the spin system is mapped onto a
solid-on-solid model (SOS)\cite{Blote}.  In the simulations, $T^*$ is
operationally defined as the temperature below which the system fails to reach
equilibrium within the simulation time.

\section{Model}

The compressible Ising antiferromagnet is described by a nearest-neighbor
spin-spin interaction which depends linearly on the distance between the
spins\cite{Chen1}.  The lattice distortions are described by a homogeneous, three component
strain field which describes the change in the three nearest-neighbor bond
lengths.  In the model studied in this paper, no fluctuations of these strain
fields are allowed.  This implies  that the strain fields are being treated within a
mean-field approximation but the spins are being treated exactly.  It has been
shown\cite{Chen1} that, in this approximation, there is a first order transition to a striped phase 
in which there are rows of aligned spins alternating in sign.  The ferromagnetic 
bonds are elongated and the anti-ferromagnetic bonds are
shortened\cite{Chen1,Leigu1}.  It has also been shown that, in this mean-field
model, there is an instability to this lattice distortion at a temperature below 
the first-order transition temperature\cite{Chen1}.

The strain fields which remove the degeneracy in the compressible model play a
role that is similar to the anisotropy of interactions\cite{Blote1} or a
staggered field which is conjugate to one of the degenerate
ground-states\cite{cdgupta}.  The crucial difference between the models is that the
strain fields are annealed variables whereas the anisotropy and the staggered
field are quenched variables.  The dynamics that we observe in the compressible
model as the glass transition is approached can, however, be rationalized on the 
basis of known results for the models with the quenched variables.  

Monte Carlo simulations were
used to study the dynamics following
instantaneous  quenches from a high-temperature disordered phase to a range of
temperatures below the ordering transition  (which was strongly
first-order) at $T_F$. Standard spin-exchange
dynamics
was extended to include
moves which attempt global changes of the shape and size of the
box\cite{Leigu1,Leithesis}.  These global changes were attempted after a complete sweep
of all the spins in the lattice.  This dynamics was adopted because the
homogeneous (global) strain fields are expected to  respond only to extensive
changes of the nearest neighbor spin correlations\cite{Chen1,Leigu1}.  There is
an ``effective'' spin model that the compressible Ising antiferromagnet can be
mapped onto.  This model involves long-range four-spin interactions. This
mapping shows the connection between the current model and the p-spin models and 
the Bernasconi model\cite{BouMez}.  The dynamical model, as defined in this
work, however, does not correspond to the dynamics of this four-spin model but
instead keeps the strain field and the spins explicitly.

\section{Spin-Spin autocorrelation functions and non-linear dynamic
susceptibility}

\begin{figure}[h]
\epsfxsize=5.4in \epsfysize=3.0in
\epsfbox{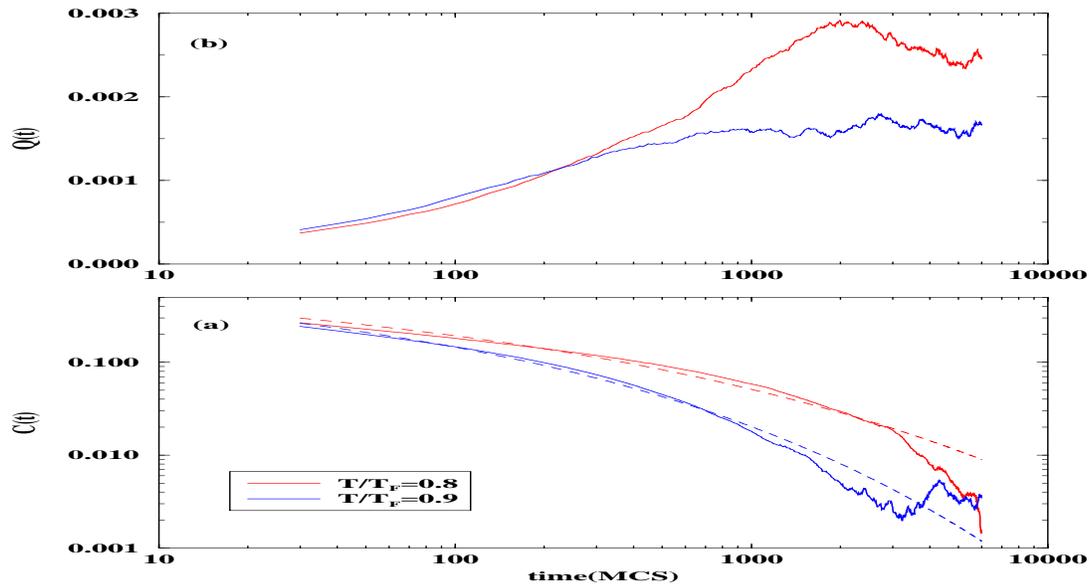}
\caption{(a)The spin-spin autocorrelation function at temperatures approaching
$T^* \simeq 0.75T_F$. The system size is 96x96. The dashed curves are
stretched-exponential fits to the data. (b) the susceptibility associated with the time-dependent overlap at the
same temperatures as shown in Fig. 1(a).  There is evidence of a peak developing as
the temperature is lowered towards $T^*$.}  
\label{fig1}
\end{figure}

As mentioned in the introduction, the system is in equilibrium above $T^*$ but
falls out of equilibrium at $T^*$.  The behavior of the spin-spin autocorrelation 
function, as $T^*$ is approached from above, is shown in Fig. \ref{fig1}.  These 
correlation functions show clear non-exponential relaxations.  The best
stretched exponential fits ($\exp(-t/\tau)^{\beta}$) are also shown in Fig. \ref{fig1}.  The
stretching parameter $\beta \simeq 0.3$ with a small dependence on temperature.
There are indications of a plateau developing at long times, however, data over
much longer times and larger systems will be needed to bear this out.  In addition to the correlation functions, Fig. 1 also shows dynamic
susceptibility associated with the fluctuations of the spin-spin correlation
function. 
This susceptibility, associated with the time-dependent
overlap\cite{Donati}, shows the same
features similar to that observed in a Lennard-Jones binary mixture\cite{Donati}.
These results suggest that the ergodicity-breaking transition at $T^*$ is akin 
to a structural glass transition even though the system being studied is a spin
system.  The hope then is that the study of this simple spin model could provide 
some insight into the structural glass transition.  It has already been shown
that a trapping model, based on the observations in this spin system, can
provide a qualitatively correct description of the observed frequency-dependent
susceptibility in structural glasses\cite{Ignatiev}. The SOS mapping provides an 
elegant way of analyzing the structures which develop in this spin model as the
glass transition is approached.  In the following, this mapping will be used to
probe the nature of the glass transition.

\section{Defects and Strings}

\begin{figure}[h]
\epsfxsize=5.4in \epsfysize=3.0in
\epsfbox{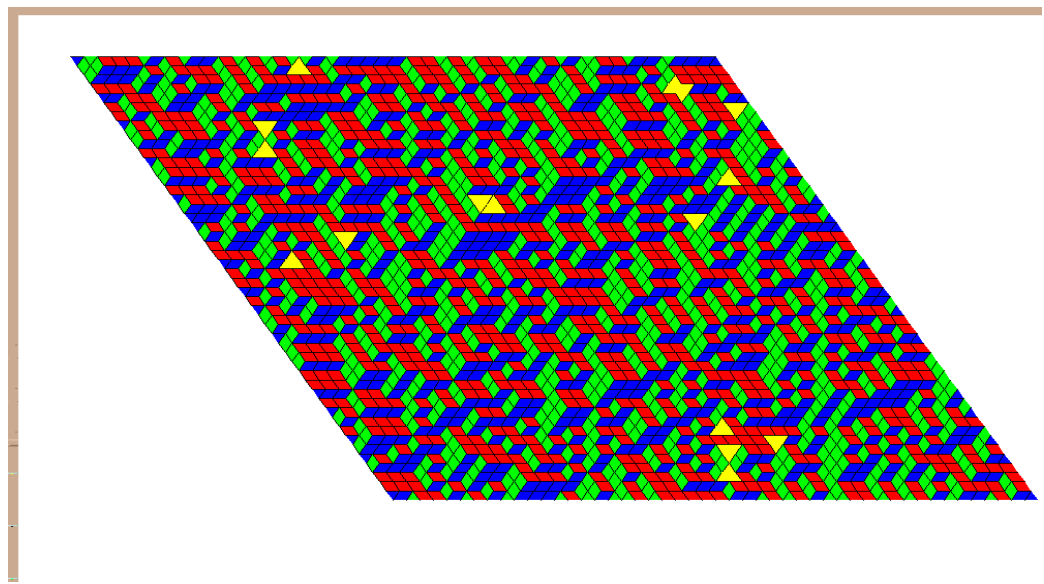}
\caption{The SOS representation of a spin configuration following an
instantaneous quench to a temperature $T^*<<T<T_F$.  The blue-red rhombi define the
```vertical'' strings ({\it cf} text).  The large yellow triangles are the local defects which
can terminate two strings}  
\label{fig2}
\end{figure}

The ground-state of the triangular-lattice Ising antiferromagnet is a critical
state and can be mapped onto the rough phase of an SOS model\cite{Blote}. In
this representation, a line is drawn between two spins which are connected by an 
anti-ferromagnetic bond\cite{Blote}.  At zero temperature, this defines a tiling of the
plane by three different types of rhombi (blue, red and green in Fig. \ref{fig2}).  At
finite temperatures, defects appear which correspond to elementary triangles
with three spins of the same sign and correspond to screw dislocations in the
SOS surface.  Fig. \ref{fig2} shows this representation for a configuration generated in
the {\it compressible} Ising model after a quench above $T^*$.  A convenient way
of representing 
the SOS surface is by strings running in the ``vertical'' direction. Choosing
one of the nearest-neighbor directions in the triangular lattice as
``horizontal'', 
these strings are defined by drawing lines connecting the middles of horizontal
edges of the rhombi.  In Fig. \ref{fig2}, the blue and red rhombi have
horizontal edges but the green do not. These strings can end at the defects
which change the number of strings by two\cite{Blote}.  The ground-state of the
pure triangular
lattice antiferromagnet corresponds to a rough SOS surface with no average tilt
and is characterized by the number of strings $N_s = 2L/3$ where $L$ is the
linear dimension of a finite lattice with periodic boundary conditions.  The
strings and defects (dislocations) provide an useful way of visualizing the
transition at $T^*$.  To understand what could be happening at $T^*$, we need to
briefly discuss the properties of the compressible Ising antiferromagnet and its 
ground-state.

The ground-state of the compressible model is the ``striped phase''\cite{Chen1}.
In the SOS picture this corresponds to a surface with a tilt and no ``vertical''
strings.  There are, obviously, three ways of defining vertical strings in the triangular 
lattice (corresponding to the three nearest neighbor directions) and the striped
phase is three-fold degenerate.  These states have only one variety of the
rhombi.  

\begin{figure}[h]
\epsfxsize=5.4in \epsfysize=3.0in
\epsfbox{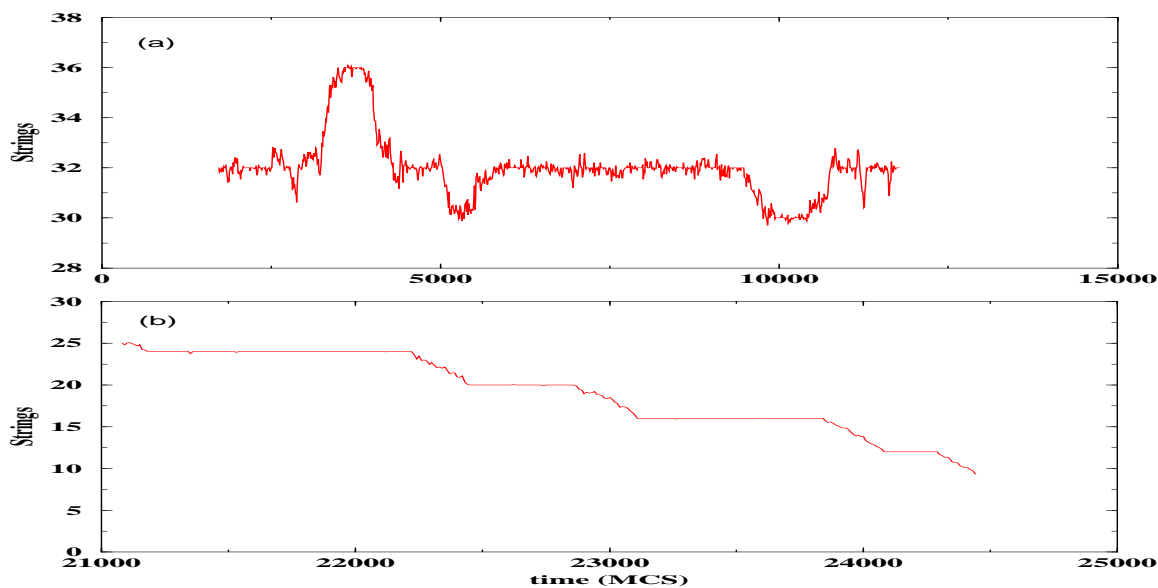}
\caption{(a) String history for temperature above $T^*$ and system size 48*48,
showing long-lived fluctuations with string density significantly different from 
$2/3$.(b) string history for
temperature well below $T^*$, for a particular initial condition.  At these
temperatures the history is extremely sensitive to initial conditions.  This
particular trajectory shows the presence of activated processes between
different string sectors.}  
\label{fig3}
\end{figure}

In terms of the SOS picture, the first-order transition at $T_F$
corresponds to a discontinuous change in the number of strings.  Strings can end
only at the defects (two strings end at a defect) and therefore the kinetics of
this transition is dominated by the interplay between defects and strings.
Similarly, one expects that the dynamics of the supercooled phase is dictated by 
the dynamics of defects and strings.  The time evolution of the number of
strings is shown in Fig. \ref{fig3} for a temperature above $T^*$ and a temperature below
$T^*$.  Above $T^*$, but close to it, the string density fluctuates around $N_s
=2L/3$ but below $T^*$ the string density is evolving towards zero in a
step-wise fashion.  The temperature $T^*$ seems to correspond to an instability
of the system towards the disappearance of strings.  It however does not seem to 
correspond to a simple spinodal since the system stays in states with a fixed
number of strings for very long times before hopping to a state with a fewer
number of strings.  This dynamics suggests a free-energy surface in this
string-number space which has multiple minima with a distribution of barrier
heights and a bias, below $T^*$, towards smaller number of strings.
This is exactly the type of picture that was used to construct the simple Langevin
model of the frequency-dependent response in glasses\cite{Ignatiev}. The order
parameter in that model corresponds to the average number of strings in the SOS
model. 

\begin{figure}[h]
\epsfxsize=5.4in \epsfysize=3.0in
\epsfbox{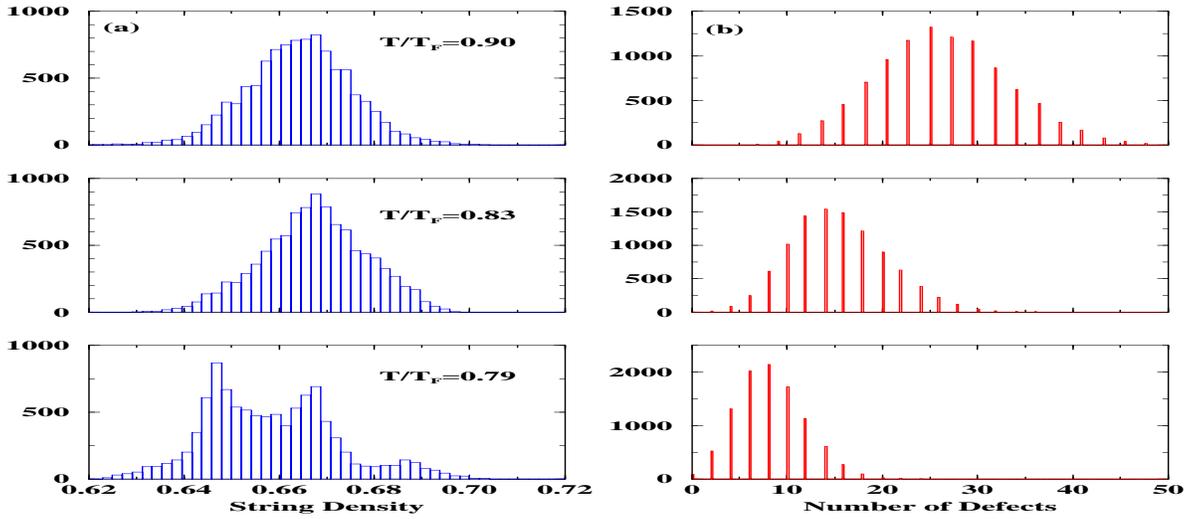}
\caption{(a) String-density  distribution at different temperatures approaching
$T^*$ in a system of size 96*96. At $T>>T^*$,  the distribution is nearly
Gaussian and peaked around $2/3$.  As $T \rightarrow T^*$, the distribution gets 
broader and shows multiple peaks which suggests that the system is finding
pathways into sectors with string density different from $2/3$.  The deviations
are both on the positive and negative side since the system fluctuates between
different orientations of the strain field and we are showing only the
``vertical'' string density as defined in Fig. 2. (b)Defect density distribution
for these same set of temperatures showing the peak narrowing and shifting
towards zero.}  
\label{fig4}
\end{figure}

Before analyzing the dynamics of the defects and strings, it is useful to look
at some static quantities above $T^*$ (where the system is in equilibrium and
static averages can be defined.  Fig. \ref{fig4} shows the histogram characterizing the
probability distribution of the average number of strings in a particular
vertical direction.  As the temperature is lowered, the
distribution develops a pronounced tail and secondary maxima.  At the same time, 
the distribution of defects changes from a nearly Gaussian distribution to a
nearly exponential one with a peak close to zero.  The combination of these two
features gives a clue to the origin of the slow dynamics.  The instability at
$T^*$ implies a tendency of the system to have fewer strings than that
characterizing the ideal supercooled state with $N_s = 2L/3$.  To achieve this
it needs to create fluctuations which have a large amplitude and last for long
times.  These string fluctuations can be created only via correlated defect
events. The type of defect events that change the number of strings by a large
number are the appearance of more than one pair of defects (defects always get
created in pairs in the dynamics that is being used in the simulations).  These
events are extremely rare and can lead to long correlation times.

\begin{figure}[h]
\epsfxsize=5.4in \epsfysize=3.0in
\epsfbox{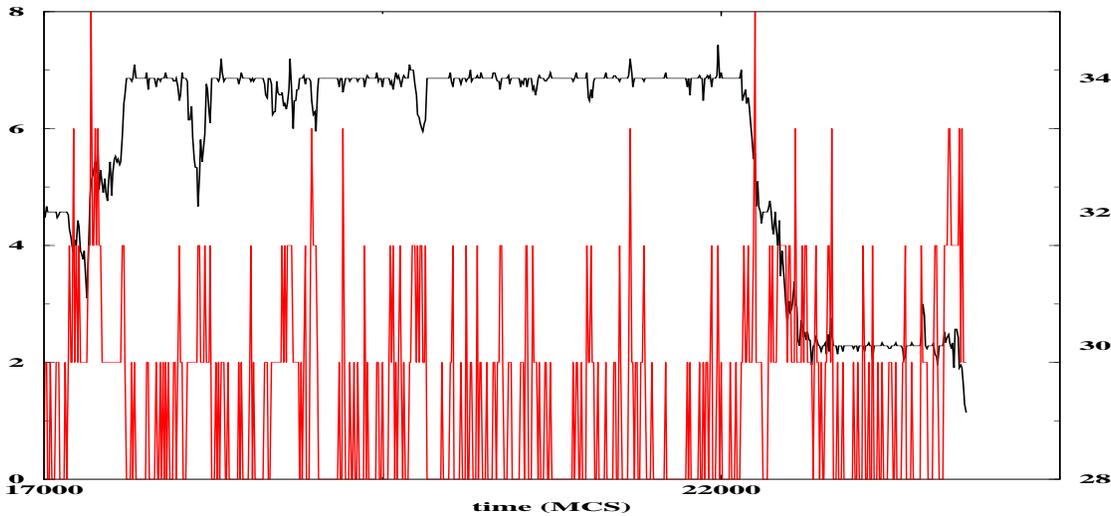}
\caption{String history and defect history at a temperature slightly above $T^*$ 
showing the type of defect events which lead to large fluctuations in the number 
of strings.  The black line shows the string history and corresponds to the
numbers on the right and the red line depicts the defect history.  The ``ideal'' 
string density in this 48x48 system is 32.}  
\label{fig5}
\end{figure}

The correlation between defect and string histories is shown in
Fig. \ref{fig5}.  To change the average number of strings in the system by two
or more, a pair of strings running the length of the system has to be created or
destroyed.  The history shown in Fig. \ref{fig5} indicates that this is not
accomplished by one pair of defects running through the system and creating or destroying strings in a ``zipper'' type action.  Instead,
the strings are created or destroyed through 
the creation of more than one defect pair in succession.  As the temperature approaches 
$T^*$, these events get rarer and the structures with significantly more (or
less) strings than the ideal supercooled state get frozen in for longer and
longer times.  The glass transition, in this frustrated spin system, seems to be 
pinned by an underlying instability (true instability in mean-field) towards a
deformation of the lattice or the spontaneous disappearance of strings.  The
dynamics approaching this transition is dictated by rare events which involve
correlated defects.  

\section{Conclusion}

In this preliminary study of correlation between defects and glassy dynamics in
a frustrated spin model, the observations suggest that defects are crucial to
understanding the glass transition.  The dynamics approaching
the transition is anomalous because of correlations between defects  and
strings.  The defects (dislocations) are local and owe their origin to the frustration in the system. 
The strings  run the length of the system and define the global
characteristics of the SOS surface and hence the ordering in the compressible
model.  In the mean-field model, there is a true instability towards the
disappearance of these strings.  In a non-mean-field model one expects to see a
pseudo instability.  To understand the slowing down of the dynamics as the
temperature approaches $T^*$, the instability temperature, it is useful to
analyze the nature of this instability a little further. The results from the
models with the quenched frustration-removing
fields\cite{Blote1,cdgupta} prove to be useful in understanding the instability
at $T^*$.  

The ground-states of the triangular lattice antiferromagnet can be classified
into sectors characterized by the string density.  In the zero-defect sector,
the free energy of the quenched models can be calculated as a function of the
string-density and the field strength\cite{Blote1,cdgupta}.  This free energy is 
minimized by a particular value of the string density for a given value of the
staggered field\cite{cdgupta} or the strength of the anisotropy\cite{Blote1}.
The analysis of Chen and Kardar\cite{Chen1} can be repeated using the language
of the strings (at least within the zero-defect sector) and indeed one finds
that the ``effective'' free energy expressed as a function of the strain fields
exhibits an
instability as the temperature is lowered.  The curvature of the effective free
energy goes to zero at the instability temperature $T^*$.  Above this
temperature there is a well-defined free-energy valley centered at zero strain
field and a string density of $2/3$. As this state loses its stability the system 
becomes free to explore  regions with finite strain.  At each realization  of the
strain field there is, however, a well-defined minimum at a definite string
-density different from $2/3$.  The instability temperature marks the point
where the system starts to see the multivalleyed nature of the free-energy
surface. The exploration of this phase space can take place 
only through activated processes which take the system from one string sector to 
another.  The process is activated since the number of strings can change only
through the creation of defects. This picture provides a qualitative
understanding of the dynamics observed in our simulations.

In continuing studies of this system, we are exploring the {\it spatial}
correlations of defects.  The spatio-temporal correlations which develop in this 
system as $T^*$ is approached would be crucial in understanding the detailed
nature of the long-lived structures and their similarity to the ones observed
in structural glass.

This work has been supported in part by the DOE grant DE-FG02-ER45495.

\newpage

\end{document}